\documentclass[pdflatex,sn-mathphys-num]{sn-jnl}

\usepackage{graphicx}%
\usepackage{multirow}%
\usepackage{amsmath,amssymb,amsfonts}%
\usepackage{amsthm}%
\usepackage{mathrsfs}%
\usepackage[title]{appendix}%
\usepackage{xcolor}%
\usepackage{textcomp}%
\usepackage{manyfoot}%
\usepackage{booktabs}%
\usepackage{algorithm}%
\usepackage{algorithmicx}%
\usepackage{algpseudocode}%
\usepackage{listings}%
\usepackage{epstopdf}%
\usepackage{svg}
\usepackage{amsmath}
\usepackage{xcolor}
\usepackage{ulem}
\usepackage{siunitx}

\begin{document}

\title[]{Cooperative Emission from Quantum Emitters in Hexagonal Boron Nitride Layers}

\author*[1,2]{\fnm{Igor} \sur{Khanonkin}} \email{ikhanonkin@technion.ac.il}
\equalcont{These authors contributed equally to this work.}

\author[2,3]{\fnm{Amir} \sur{Sivan}}
\equalcont{These authors contributed equally to this work.}

\author[1,4]{\fnm{Le} \sur{Liu}}

\author[1]{\fnm{Johannes} \sur{Eberle}}

\author[5]{\fnm{Kenji} \sur{Watanabe}}

\author[6]{\fnm{Takashi} \sur{Taniguchi}}

\author[2,3]{\fnm{Gadi} \sur{Eisenstein}}

\author[2,3]{\fnm{Meir} \sur{Orenstein}}

\affil[1]{\orgdiv{Institute for Quantum Electronics}, \orgname{ETH Zurich}, \city{Zurich 8093}, \country{Switzerland}}

\affil[2]{\orgdiv{Faculty of Electrical and Computer Engineering}, \orgname{Technion - Israel Institute of Technology}, \city{Haifa 3200003}, \country{Israel}}

\affil[3]{\orgdiv{Helen Diller Quantum Center}, \orgname{Technion - Israel Institute of Technology}, \city{Haifa 3200003}, \country{Israel}}

\affil[4]{\orgdiv{School of Electrical and Electronic Engineering}, \orgname{Nanyang Technological University}, \city{Singapore 637616}, \country{Singapore}}

\affil[5]{\orgdiv{Research Center for Electronic and Optical Materials, National Institute for Materials Science}, \orgname{NIMS}, \city{Tsukuba 305-0868}, \country{Japan}}

\affil[6]{\orgdiv{Research Center for Materials Nanoarchitectonics, National Institute for Materials Science}, \orgname{NIMS}, \city{Tsukuba 305-0868}, \country{Japan}}

\abstract{
Collective light emission from many-body quantum systems is a cornerstone of quantum optics, yet its realization in solid-state platforms operating under ambient conditions remains highly challenging. Large-bandgap van der Waals materials such as hexagonal boron nitride (hBN) host stable room-temperature single-photon emitters with narrow linewidths across a broad spectral range. However, cooperative radiative effects in this system have not been previously explored. Here we demonstrate collective emission from quantum-emitter ensembles in hBN layers when the emitters are nearly indistinguishable and positioned within a sub-wavelength proximity. Using confocal microscopy and a Hanbury Brown–Twiss (HBT) configuration, we identify both isolated emitters and ensembles activated by localized electron-beam irradiation. Time-resolved photoluminescence measurements reveal a superlinear intensity enhancement and a pronounced acceleration of the radiative decay in tightly confined ensembles, with lifetimes approaching the temporal resolution of our experimental system (about \qty{500}{ps}), compared to approximately \qty{1.85}{ns} for single emitters or large, spatially extended ensembles. Complementary second-order photon-correlation measurements exhibit subpoissoninan antidip consistent with emission from a few indistinguishable emitters. The simultaneous observation of lifetime shortening and enhanced emission provides direct evidence of cooperative emission at room temperature—achieved without optical cavities or cryogenic cooling. These results establish optically active defect ensembles in hBN as a scalable solid-state platform for engineered collective quantum optics in two-dimensional materials, opening avenues toward ultrabright superradiant light sources and nonclassical photonic states for quantum technologies.
}


\maketitle

\section*{Introduction}\label{sec1}

The collective behaviour of quantum emitters has long fascinated both quantum optics and condensed matter physicists. A paradigmatic example is superradiance, first introduced by Dicke in 1954, where an ensemble of $N$ identical two-level systems radiates cooperatively into a common electromagnetic mode~\cite{Dicke1954}. In this regime, the indistinguishability of the emitters leads to the formation of symmetric Dicke states constituting a $N+1$ levels degenerate energy ladder, which upon excitation spontaneously emit photons with intensity that scales as $I \propto N^2$ and radiative lifetimes that are shortened by a factor of up to $1/N$. Such cooperative enhancement of spontaneous emission not only reveals fundamental many-body dynamics but also offers pathways to ultrabright, ultrafast, and coherent quantum light sources.

Superradiance occurs when an excited but incoherent ensemble of emitters spontaneously builds up coherence through quantum fluctuations, leading to an amplified and accelerated emission of photons resulting from the constructive quantum interference of the ensemble evolution paths ~\cite{Bonifacio1975,Gross1982}. Despite its conceptual simplicity, this phenomenon has garnered significant attention over the years, and its theory has been generalized beyond the fundamental Dicke model to include geometry and matter ~\cite{john1995localization, pustovit2010plasmon, sivan2019enhanced, asenjo2017exponential}, coupling to cavities and photonic structures ~\cite{pineiro2022emergent, sundar2024squeezing, sheremet2023waveguide, solano2017super}, emitters with complex structures ~\cite{crubellier1985superradiance, sivan2025adding} and more.

Experimentally, superradiance has been demonstrated in diverse platforms: atomic gases and Bose--Einstein condensates~\cite{Skribanowitz1973,Inouye1999}, cold atoms in optical cavities~\cite{Bohnet2012,Meiser2010}, self-assembled quantum dots~\cite{Scheibner2007}, high density of nitrogen-vacancy (NV) centres \cite{bradac2017room} and two-coupled NV emitters embedded in diamond nanopillars \cite{dastidar2024signatures}, perovskite nanocrystal superlattices~\cite{Raino2018, Levy2025}, coherent excitons van der Waals heterostructures~\cite{Haider2021}, and semiconductor heterostructures~\cite{laurent2015superradiant}. These results underscore the universality of cooperative spontaneous emission but also highlight its fragility in solid-state media, where inhomogeneous broadening, phonon coupling, and disorder suppress collective coherence. Importantly, room-temperature demonstrations in defect-based emitters remain rare, and no clear evidence has yet been established in two-dimensional quantum materials.

Hexagonal boron nitride (hBN) provides a unique opportunity to address this challenge. As a wide-bandgap van der Waals crystal, hBN hosts optically stable single-photon emitters that operate robustly from cryogenic to ambient conditions~\cite{ccakan2025quantum,huang2022engineering,su2024fundamentals,kubanek2022coherent}. Among them, the so-called blue centers (B-centers) stand out. Recent first-principles studies have identified these emitters as carbon-chain tetramers incorporated into the hBN lattice~\cite{Maciaszek2024,DefectsUltraThin2025}. These defects exhibit a sharp zero-phonon line at $436~\text{nm}$, radiative lifetimes of $1.8-1.9~\text{ns}$, and excellent photostability—well matching the requirements for applications in quantum sensing based on ultra-thin hBN~\cite{DefectsUltraThin2025}. B-centers can be considered close to indistinguishable \cite{gerard2025resonance}, which means that emitters share the same electronic transition and emission spectrum within the homogeneous linewidth, such that their radiative pathways interfere coherently on the timescale of spontaneous emission. Moreover, because they can be reproducibly generated at selected positions by electron irradiation~\cite{Fournier2021}, they serve as ideal candidates for exploring cooperative emission at the nanoscale.

Here we demonstrate superradiant emission from ensembles of B-center defects in hBN layers at room temperature, which are effectively two-levels emitters that are nearly indistinguishable and positioned within sub-wavelength proximity. Realistically, the radiative lifetime and intensity of the emitter ensemble will not scale as $1/N$ and $N^2$ respectively as in the theoretical Dicke superradiance model, since the detailed ensemble geometry and relative emitter dipole orientations, which are hard to control in the defect activation process (see \nameref{sec11}), play a crucial role in the dynamics of cooperative emission \cite{sivan2019enhanced}. However, the signature of collective emission is clearly evident in our experiment. By comparing a single B-center, characterized by a radiative lifetime of $\tau \approx 1.85,\mathrm{ns}$, with localized ensembles containing estimated $N=1,2,3$ and $4$ emitters, we resolve a systematic and monotonic acceleration of the radiative decay, reaching values below the temporal resolution of our experimental system ($\approx \qty{500}{ps}$), accompanied by a superlinear enhancement of the emission intensity. Moreover, we emphasize the quantum nature of our observations which is manifested as subpoissonian bunched second-order photon correlation measurements  \cite{mandel1995optical}. The immediate onset and reproducibility of the lifetime shortening establish these few-emitter defect ensembles as a solid-state room-temperature realization of collective emission in a two-dimensional material platform.

Beyond fundamental significance, our results point to compelling applications. First, hBN emitters already support room-temperature electroluminescence~\cite{Fournier2020}, suggesting the feasibility of a defect-based superradiant laser in a two-dimensional material platform, a regime so far realized only in ultracold atomic systems \cite{bohnet2012steady}. Unlike conventional semiconductor lasers, such devices would be stabilized by atomic coherence rather than cavity quality factors, promising spectral purity and high brightness. Second, cooperative emission can profoundly alter photon statistics. Recent theoretical studies of superradiant photonic states have revealed non-Gaussian Wigner functions \cite{sivan2025adding}, hallmarks of nonclassical light. Importantly, the concept of quantum non-Gaussianity has been formalized as a property of states that cannot be expressed as mixtures of Gaussian states, even when their Wigner functions remain positive~\cite{Filip2011criterion}, and this has been experimentally verified using heralded single-photon sources~\cite{Jezek2011exp}. Superradiance from B-center ensembles in hBN may therefore provide a solid-state pathway toward bright, room-temperature quantum non-Gaussian light, with potential applications in quantum communication, computation, and metrology.

\section*{Results and Discussion}\label{sec2}

\begin{figure}[ht]
    \centering
    \includegraphics[width=1\linewidth]{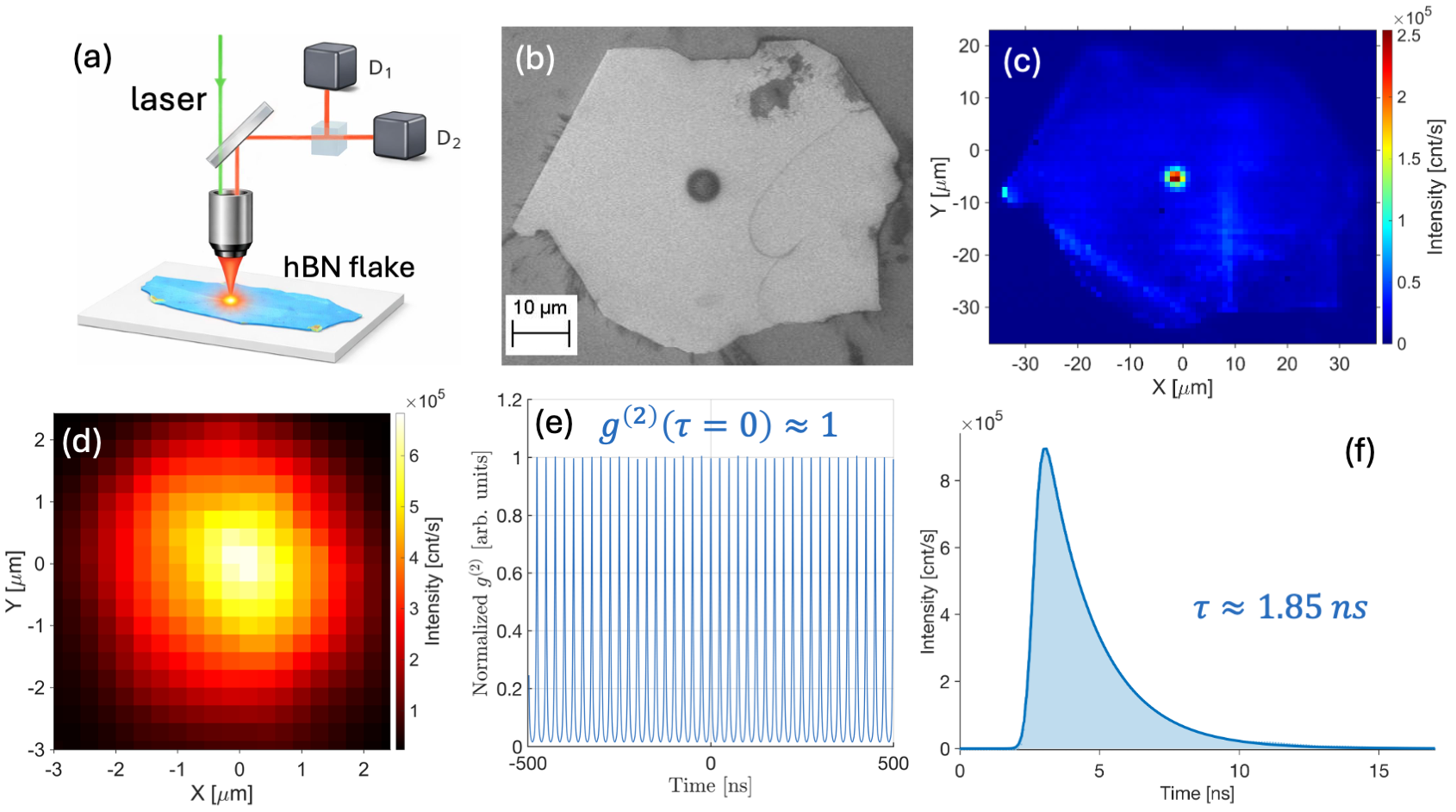}
    \caption{
    \textbf{Large and uncorrelated ensemble of quantum emitters.}
    (a) Schematic of the photoluminescence (PL) measurement from hBN optical active defects in a Hanbury Brown–Twiss (HBT) configuration. 
    (b) SEM image of the hBN flake after a strong localized irradiation, showing an activated defect site (black spot; irradiation time $\sim 15$ min).
    (c) wide-field confocal PL scan identifying an irradiated region as a defect ensemble.
    (d) Zoom-in confocal PL map revealing the quantum-emitter ensembles.
    (e) Normalized second-order photon correlation $g^{(2)}(\tau)$ for the ensemble, showing $g^{(2)}(0) \approx 1$ corresponding to uncorrelated ensemble of quantum emitters.
    (f) Time-resolved PL traces for the typical large and uncorrelated ensemble, showing mono-exponential lifetimes of $1.85$ ns.
   }
    \label{Fig_1}
\end{figure}

 A schematic of the photoluminescence (PL) measurement is shown in Fig. \ref{Fig_1} (a). To assess whether cooperative emission persists in larger defect ensembles, we examined regions subjected to prolonged irradiation (electron dose of $4.5 \times 10^{13}$) that generate high densities of B-centers. SEM imaging (Fig. \ref{Fig_1} (b)) and PL mapping (Fig. \ref{Fig_1} (c,d)) reveal extended, spatially diffuse emission consistent with a large ensemble rather than a few tightly clustered emitters. Time-resolved PL measurements from these regions show purely mono-exponential decays with lifetimes of $1.85\,\text{ns}$, identical to those of isolated single emitters within measurement uncertainty and inhomogeneity of B-centers. No lifetime reduction, intensity enhancement, or photon-correlation signatures associated with superradiance are observed. These results affirm that in large, spatially extended ensembles the radiative evolution paths of the ensemble interfere destructively and cooperative emission is therefore not exhibited. 

\begin{figure}[ht]
    \centering
    \includegraphics[width=1\linewidth]{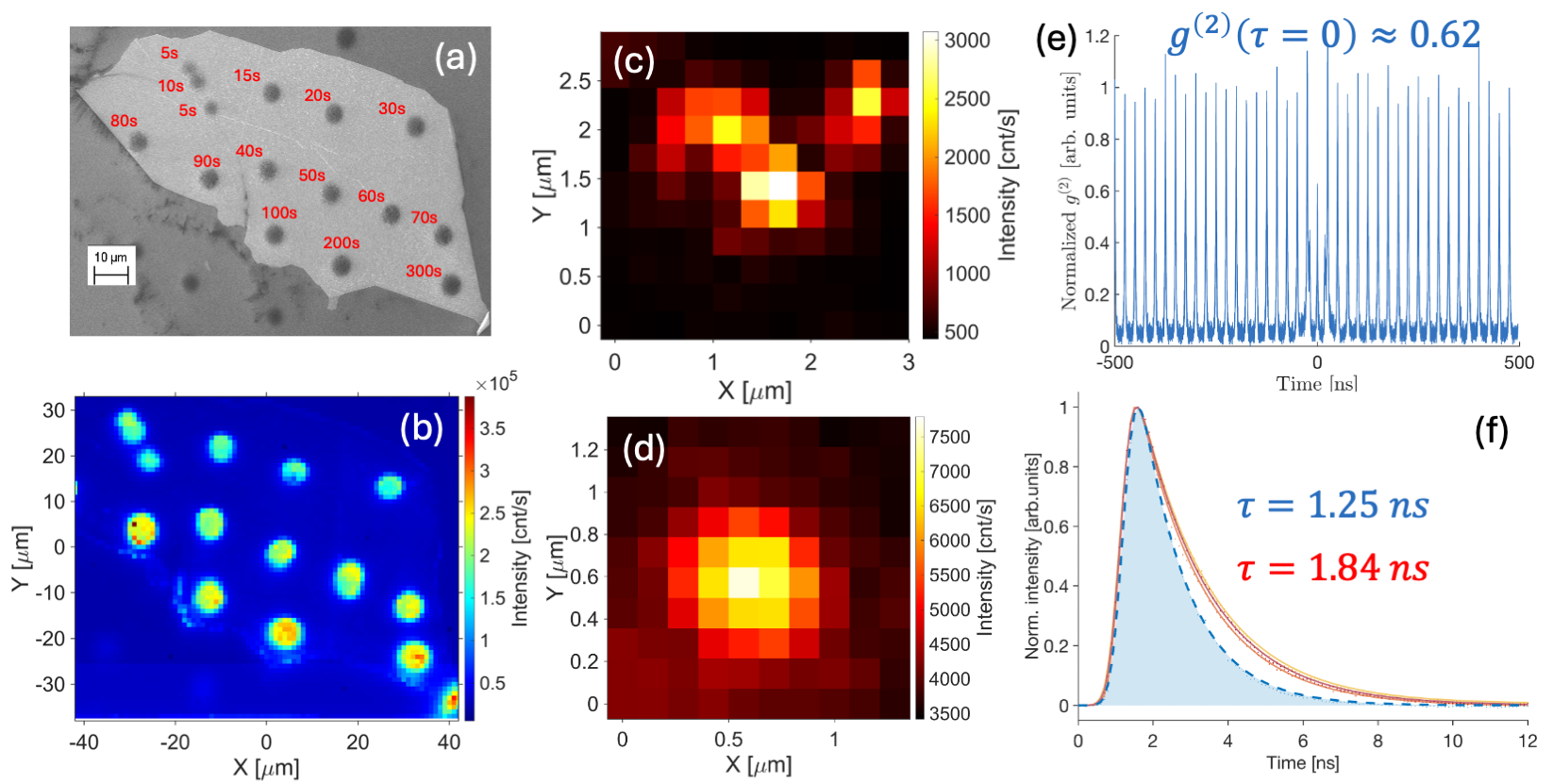}
    \caption{\textbf{Separated versus spatially localized emitters: non-cooperative and superradiant regimes.} 
    (a) SEM image of the hBN flake after multiple localized irradiations, showing activated defect sites (black spots; irradiation times 5--300 s). 
    (b) wide-field PL scan identifying each irradiated spot as a defect ensemble. 
    Conceptual comparison of emitters placed far apart (c), resulting in non-cooperative emission, versus (d) two emitters positioned within a sub-wavelength distance, enabling cooperative emission. Confocal PL map showing the spatial arrangement of the emitters. (e) Second-order photon-correlation measurement revealing a subpossonian bunching antidip of $g^{(2)}(0)\approx 0.62$, consistent with a two-emitter ensemble. (f) Time-resolved PL traces showing an accelerated decay of $1.25\,\text{ns}$ for the tightly spaced (superradiant) emitters compared to $1.84\,\text{ns}$ for the spatially separated (non-superradiant) emitters.}
    \label{Fig_2}
\end{figure}

To avoid destructive interference between a large number of random radiative evolution paths, and demonstrate collective emission in hBN, we focus on small and tightly confined ensembles comprising only a few emitters. Figure \ref{Fig_2} summarizes the identification and characterization of individual and few-emitter B-center ensembles in layered hBN. Localized electron-beam irradiation is used to generate optically active defect sites at predefined positions within the flake, as evident from the SEM image in Fig.\ref{Fig_1} (a), where each dark spot corresponds to an activated region with irradiation times ranging from 5 to 300~s. A wide-field PL scan (Fig.\ref{Fig_2} (b)) confirms that every irradiated location hosts a bright defect ensemble.

We now demonstrate the role of indistinguishability in enabling collective emission. A confocal two-dimensional PL map of the fabricated structures is shown in Fig.\ref{Fig_2} (a,b), revealing two regions with comparable overall brightness but distinct spatial extent. One region contains a tightly clustered group of emitters, whereas the second consists of three emitters separated by several wavelengths. Time-resolved PL traces (Fig.\ref{Fig_2} (d)) clearly differentiate the two regimes. The spatially separated emitters display a radiative lifetime of $1.84\,\text{ns}$, matching that of non-interacting B-centers. In contrast, the closely spaced emitters exhibit a significantly accelerated decay with a lifetime of $1.25\,\text{ns}$, consistent with an enhancement by almost 50\% of the radiative decay rate with respect to a single B-center emitter, and an intensity enhancement factor of approximately $2.5$. We estimate from our measured $g^{(2)}(0)$ value of $\sim0.62$ and the aforementioned emission enhancement factors that the confined ensemble consists of two emitters. This estimation is derived from the lower bound of $g^{(2)}(0)$ of the case of $N$ independent emitters, namely $g^{(2)}_{ind}(0)=1-1/N$. If cooperativity exists, a bunching antidip forms at $\tau=0$ resulting in $g^{(2)}(0)>g^{(2)}_{ind}(0)$ \cite{koong2022coherence, machielse2019quantum, cygorek2023signatures}. The measured value of $g^{(2)}(0)\approx0.62$ suggests a small bunching with respect to the lower bound value of $g^{(2)}_{ind}(0)=0.5$ implying small cooperativity - consistent with our measured sub-linear and sub-quadratic rate and intensity enhancement factors with respect to $N$. This illustrates the conceptual distinction: when several B-centers are positioned within a sub-wavelength volume (thereby making them indistinguishable) - a collective emission is observed. In contrast, when the emitters are spaced farther apart, they radiate independently, resulting in non-cooperative behavior.

\begin{figure}[ht]
    \centering
    \includegraphics[width=1\linewidth]{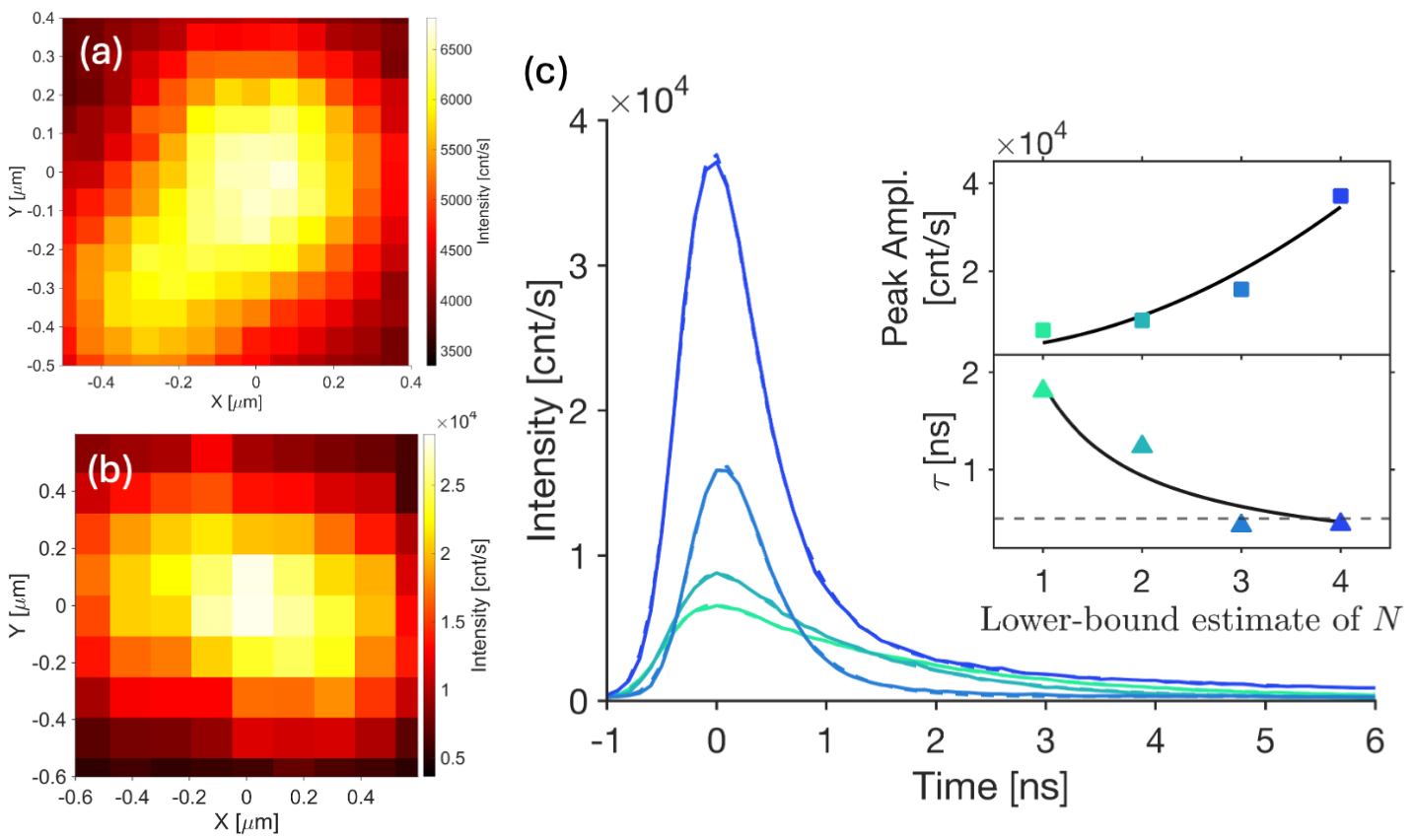}
    \caption{\textbf{Superradiant scaling with ensemble size.}
(a,b) Two-dimensional PL maps of a tightly confined emitter ensemble within a diffraction-limited area, exhibiting a pronounced super-linear enhancement of the PL intensity.
(c) Non-normalized time-resolved PL traces recorded under identical excitation and detection conditions for ensembles containing a lower-bound estimate of $N=1$, $2$, $3$ and $4$ emitters. The inset shows the extracted peak PL amplitude and radiative lifetime $\tau$, obtained from mono-exponential fits, as a function of $N$. For $N=4$, the decay is better described by a bi-exponential fit, and the fast component is used to extract the effective radiative lifetime. Uncertainties in extracted values of $\tau$ do not exceed \qty{20}{ps}. Ensembles with higher estimated number of emitters exhibit pronounced lifetimes shortening that approach and fall below the temporal resolution of the measurement system (\qty{500}{ps}). Black curves are shown to demonstrate the trend.} 
    \label{Fig_4}
\end{figure}

Figure~\ref{Fig_4} summarizes the scaling behavior of collective emission as a function of the number of emitters within a tightly confined ensemble. Figures~\ref{Fig_4}(a,b) show two-dimensional PL maps of a representative ensemble, revealing a highly localized emission region accompanied by a pronounced shortening of the radiative lifetime. Time-resolved PL traces recorded under identical excitation and collection conditions for ensembles containing an estimated lower-bound number of emitters $N =1, 2, 3$ and $4$ are presented in Fig.~\ref{Fig_4}(c). The lower bound is dictated by the rate enhancement factors. Note that for $N=3$ and $N=4$ the measured lifetime approaches the temporal resolution of our measured system of about \qty{500}{ps}. The extracted peak PL amplitude and radiative lifetime $\tau$ are summarized in the inset of Fig.~\ref{Fig_4}(c), with black line curves showing the trend. While mono-exponential fits adequately describe the decay dynamics for smaller ensembles, the $N = 4$ trace is better captured by a bi-exponential model, reflecting the coexistence of collective and residual non-collective or dark state decay channels; in this case, the fast component is used to define the effective radiative lifetime. Together, these observations provide clear experimental evidence of cooperative emission arising from coherent coupling within the emitter ensemble.

\section*{Conclusion}

In summary, we have demonstrated cooperative spontaneous emission from ensembles of optically active defects in hexagonal boron nitride at room temperature. By identifying small groups of nearly indistinguishable B-center emitters within a sub-wavelength volume, we observe a pronounced acceleration of the radiative decay accompanied by enhanced emission intensity. Time-resolved photoluminescence measurements of tightly confined ensembles exhibit significant radiative lifetime shortening, reaching values close to the temporal resolution of our detection system for some of the ensembles, and PL enhancement that scales super-linearly with the number of emitters. Concurrent second-order photon-correlation measurements confirm emission from few B-center emitters and establish the cooperative nature of the observed dynamics.

Importantly, because all emitters reside within the same hBN flake and optical environment, and because spatially separated ensembles with comparable brightness do not exhibit lifetime shortening, the observed decay acceleration cannot be attributed to variations in the local density of optical states or Purcell-type effects. This conclusively links the lifetime reduction to cooperative radiative coupling enabled by sub-wavelength confinement and emitter indistinguishability.

Our results establish hexagonal boron nitride as a unique solid-state platform in which collective emission emerges robustly under ambient conditions, without the need for optical cavities or cryogenic cooling. The generation of spatially confined defect ensembles using electron-beam irradiation enables direct access to the crossover between independent and cooperative emission regimes in a two-dimensional material.

Beyond their fundamental significance, these findings open promising directions toward ultrabright and ultrafast quantum light sources operating at room temperature, including defect-based superradiant lasers and electrically driven platforms. More broadly, this system offers a versatile testbed for exploring many-body quantum optics, engineered dissipation, and collective quantum states in two-dimensional materials, with potential applications in quantum communication, sensing, and integrated quantum photonics. Taken together, our work demonstrates that defect-based quantum emitters in van der Waals materials can serve not only as isolated single-photon sources but also as scalable building blocks for engineered cooperative light–matter interactions at the nanoscale.

\section*{Methods}\label{sec11}

\subsection*{Sample preparation}
The high-pressure high-temperature hBN crystal used in this study was sourced from the National Institute for Materials Science (NIMS). Thin layers of hBN were prepared by mechanical exfoliation onto $\mathrm{SiO_2}$ (\qty{285}{\nano\meter})/Si substrates. The hBN thickness was determined using optical reflection contrast. The samples were subsequently irradiated with electron beams using a Zeiss ULTRA~55~Plus system. The acceleration voltage is \qty{15}{\kilo\volt} and the current is \qty{8}{\nano\ampere}. The diameter of the irradiation spot is smaller than $\qty{10}{\nano\meter}$, which in principle allows for the creation of localized B centers. However, the focused electron beam leads to layer etching of the hBN, as discussed in Ref.~\cite{DefectsUltraThin2025}. Therefore, the B center ensembles investigated in the present work were located in regions of the hBN flake spatially separated from the irradiation spot, where the hBN lattice is not damaged by the electron beam. The studied B-center ensembles---exhibiting different numbers of emitters and varying relative separations---were not deterministically defined by the SEM irradiation, but instead emerged from a stochastic defect-formation process driven by electron scattering events in the hBN/SiO$_2$.

\subsection*{Experimental setup}
Time-resolved photoluminescence measurements were performed using 56~ps pulsed diode laser (PicoQuant) with repetition rate of 40 MHz, tuned to the B-center zero-phonon line at $\sim 440\,\text{nm}$. The excitation beam was focused onto exfoliated hBN on $\mathrm{SiO_2/Si}$ using a high–numerical-aperture objective (NA = 0.95). The sample was mounted on a piezoelectric XYZ stage with nanometer spatial resolution for precise confocal scanning. Emitted photoluminescence was collected through the same objective, spectrally filtered to isolate the phonon-sideband emission at $475 \pm 5\,\text{nm}$, and sent to a 50:50 beam splitter. The two output channels were fiber-coupled to single-photon avalanche diodes (SPADs, Excelitas) for lifetime and $g^{(2)}(\tau)$ measurements. Photon arrival times were recorded by a Swabian Time Tagger Ultra for reconstructing decay dynamics and correlation functions.

\backmatter

\bmhead{Acknowledgements}
We are grateful to Ata\c{c} Imamo\u{g}lu (ETH Zurich) for his guidance and encouragement throughout this study. We thank Shai Levy (Technion) and Maria Chekhova (Max Planck Institute for the Science of Light) for insightful suggestions and valuable discussions. I.K. acknowledges the financial assistance of the Rothschild Post-Doctoral Fellowship from Yad HaNadiv, and from the Ministry of Innovation, Science and Technology of Israel. A.S. acknowledges financial support from a partial fellowship from the Helen Diller Quantum Center of the Technion - Israel institute of Technology. K.W. and T.T. acknowledge support from the JSPS KAKENHI (Grant Numbers 21H05233 and 23H02052), the CREST (JPMJCR24A5), JST and World Premier International Research Center Initiative (WPI), MEXT, Japan.

\bibliography{sn-bibliography}

\end{document}